\documentstyle[preprint,prb,aps,amssym]{revtex}

\newcommand{\bmq}{{\mbox{\boldmath $q$}}}

\begin{document}
\preprint {WIS-99/12 Mar-DPP}
\draft

\date{\today}
\title{Theoretical aspects of the CEBAF 89-009 experiment on inclusive
scattering of 4.05 GeV electrons from nuclei}
\author{A.S. Rinat and M.F. Taragin}
\address{Department of Particle Physics, Weizmann Institute of Science,
         Rehovot 7616, Israel}

\maketitle
\begin{abstract}

We  compare  recent  CEBAF  data  on inclusive  electron
scattering  of  4.05  GeV
electrons  on  nuclei  with  predictions, based  on  a  relation  between
structure  functions (SF)  of  a  nucleus, a  nucleon  and  a nucleus  of
point-nucleons.   The  latter  contains  nuclear  dynamics,  e.g.  binary
collision  contributions  in  addition  to  the  asymptotic  limit.   The
agreement with data  is good, except in  low-intensity regions.  Computed
ternary collision contributions appear too  small for an explanation.  We
perform scaling analyses in Gurvitz's scaling variable and found that for
$y_G \gtrless 0$, ratios of scaling  functions for pairs of nuclei differ
by  less than  15-20\%  from 1. Scaling functions for $\langle
y_G\rangle <0$  are, for  increasing $Q^2$, shown  to approach  a plateau
from above.  We  observe only weak $Q^2$-dependence in FSI,  which in the
relevant kinematic region  is ascribed  to the diffractive nature of $NN$
amplitudes  appearing in  FSI.  This  renders it  difficult to  separate
asymptotic from FSI parts and  seriously hampers the extraction of $n(p)$
from scaling analyses in a model-independent fashion.

\end{abstract}

\newpage
In the following we discuss some aspects of inclusive scattering of
high-energy electrons from nuclear targets with the following
cross section per nucleon
\begin{eqnarray}
\frac{d^2\sigma_{eA}(E;\theta,\nu)/A}{d\Omega\,d\nu}
=\frac{2}{A}
\sigma_M(E;\theta,\nu)\bigg\lbrack\frac {xM^2}{Q^2}F_2^A(x,Q^2)+
{\rm tan}^2(\theta/2)F_1^A(x,Q^2)\bigg\rbrack
\label{a1}
\end{eqnarray}
The inclusive,
as well as the Mott cross for point-nucleons $\sigma_M$, are
measured  as functions  of  the  beam energy  $E$,  the scattering  angle
$\theta$ and the energy loss $\nu$.  The above $F_{1,2}^A(x,Q^2)$ are two
nuclear  structure functions (SF)  which account for the  inclusive
scattering  of
unpolarized electrons  from randomly  oriented targets.  These  depend on
the squared  4-momentum $Q^2=\bmq^2-\nu^2$ and the  Bjorken variable $x$,
corresponding to the nucleon mass $M$ with range $0\le Q^2/2M\nu\le A$.

We concentrate  on the  recent CEBAF  $E$=4.05 GeV  inclusive scattering
data on nuclear  targets, notably Fe \cite{arr}.  These cover  a range of
squared   4-momenta    transfer   $Q^2=q^2-\nu^2,\,\,1\lesssim   Q^2({\rm
GeV}^2)\lesssim 7$ and Bjorken  variable $x=Q^2/2M\nu,\,\, 0.20\lesssim x
\lesssim 4.2$, and vastly extend the angular and energy-loss ranges of
the older NA3 SLAC \cite{day} and related experiments \cite{brad,arr1}.

Predictions  for  the  above  CEBAF   experiment have  been  made  before
\cite{rt2}. These were based on  a model which connects the structure
functions (SF) of  a nucleus  $F_k^A$ and $F_k^N$ of  a nucleon \cite{gr}
\begin{eqnarray}
F_k^A(x,Q^2)=\int_0^A  \frac{dz}{z^{2-k}}f^{PN}(z,Q^2)
F_k^N   \bigg    (\frac{x}{z},Q^2\bigg   )
\label{a2}
\end{eqnarray}
A calculation of  the above nuclear SF requires two  input elements.  The
first one is $F_k^N\equiv F_k^{<p,n>}$, properly weighted with proton and
neutron numbers in the nuclear target.  The $F^N$ contain  nucleon
elastic (NE) part  in terms of the standard static  electric and magnetic
form  factors $G_E^N,G_M^N$,  and nucleon  inelastic (NI) parts which are
available  in parametrized  and tabulated  forms \cite{bod}.   The second
element   is  the   structure  function   $f^{PN}$  for   a  nucleus   of
point-nucleons,  which  accounts for  the  nuclear  dynamics in  $F_k^A$.
Calculations employ  a relativistic  extension of  the non-relativistic
(NR) Gersch-Rodriguez-Smith  (GRS) series in $1/q\,\,\,$  \cite{grs}.
The  latter   contains  an   asymptotic  limit   (AL),  related   to  the
single-nucleon momentum distribution $n(p)$, and Final State Interactions
(FSI) which  are dominated  by binary  collisions between  the knocked-on
nucleon and a nucleon from the core.

Expressions, similar to (\ref{a2}), have previously been  suggested with
light-cone    momentum   fractions    instead   of    Bjorken   variables
\cite{aj,jaffe}.  The two coincide  for $Q^2\to\infty$ and Eq. (\ref{a2})
conjectures the approximate equality to
hold for $Q^2$, in  excess of some $Q_c^2$.

The model  incorporated in (2) locates weak $A$-dependence
of  $F_k^A(x,Q^2)$ only  in  the  neutron excess  $\delta  N/2A$, and  in
$f^{PN}\,\,$ \cite{rt2}, thus
\begin{eqnarray}
F_k^A(x,Q^2)\approx F_k(x,Q^2) +{\cal O}(1/A);\,\,\,A \gtrsim 12,
\label{a3}
\end{eqnarray}
with the above  restriction on $A$, due to the  neglect of nucleon recoil
in calculations of $f^{PN}$ in (\ref{a2}).   In Fig. 1 we compare Fe data
\cite{arr} and the above predictions \cite{rt2} and conclude:

i) There is good agreement  in the deep-inelastic region $\nu\ge Q^2/2M$,
and  satisfactory  correspondence  on  the   immediate  NE  side  of  the
quasi-elastic peak (QEP), $x\gtrsim 1$.

ii) Data are  for given $\theta$, and  not for fixed $Q^2$,  but from the
fact  that $Q^2$  increases with  $\theta$, and  for given  $\theta$ with
decreasing $\nu$,  one observes  that discrepancies grow  with decreasing
$\theta$, i.e. for decreasing $Q^2$.   For $Q^2\lesssim 1.6$ GeV$^2$ this
may in  part be due to  variation in some ill-determined  $NN$ scattering
parameters,  which  enter the  FSI  calculations.   However, it  is  more
likely,   that    $Q^2\lesssim   Q^2_c(x,\theta)$,   below    which   the
representation (\ref{a2}) becomes progressively flawed.

iii) For all $\theta$, computed cross sections overestimate the data by a
factor up to 2-3 for the lowest energy losses $\nu$, where cross sections
have dropped orders of magnitude from their maximum.

In  an attempt to understand the above  discrepancy  we worked out  more
complex FSI contributions,
generated by ternary  collisions (TC).  These are on general
grounds expected to effect low-intensity  regions far from the QEP, where
they  compete with  both the  AL and  FSI, due  to BC.  Their small,  but
noticible effect has recently been  established in the response of liquid
$^4$He  $\,$\cite{rt3}.  We  have used  a relativistic  extension of  the
above TC contribution and found that these indeed contribute in the above
kinematic regions, but only with insignificant weight.

At  this point  we recall  that it  is not  at all  clear that  the above
is a real discrepancy. Using different acceptable  $n(p)$ produce results
which range over the area of the above mentioned local discrepancies (see
Figs. 5,6 in Ref. \onlinecite{rt2}).

The NA3 experiment \cite{day} has also been analysed by means of versions
of the  Plane Wave Impulse  Approximation (PWIA)  in terms of  a spectral
function,     occasionally     supplemented     by     additional     FSI
\cite{om,ciof1,ciof2,oset}.   We recall  attempts by  Benhar $et\,al$  to
ascribe  to  color transparency  a  desired  lowering of  cross  sections
\cite{om}.  Conventional $2p-1h$  FSI on the PWIA  apparently produce the
same effect \cite{ciof3}.

A  comparison with  the  above mentioned  generalized  GRS results  shows
excellent agreement  in inelastic regions.  For  decreasing energy losses
Ciofi and Simula somewhat underestimate intensities \cite{ciof3}, whereas
our approach overestimates them. In fact, one wonders how two entirely
different series,  GRS and  IA tend to  produce comparable  answers.  The
explanation is  a recent demonstration  for NR dynamics of  a surprising,
order-by-order correspondence in $1/q$ \cite{rj}.

Next we  turn to a few  selected scaling analyses, previously  applied to
the NA3 data $\,\,$\cite{rt1,arr1}. We had used there a relativistic
West-GRS scaling variable suggested by Gurvitz \cite{sag} $y_G$,
related to $x$  by $y_G\approx (M\nu/q)(1-x)$.
We investigated  ratios of inclusive cross sections
\begin{eqnarray}
\xi^{A_1,A_2}=\bigg (\frac{d^2\sigma^{eA_1}}{A_1}\bigg ) \bigg /
\bigg (\frac {d^2\sigma^{eA_2}}{A_2}\bigg )
\label{a4}
\end{eqnarray}
and  in particular  $\xi^{A,<N>}(y_G<0,q)$  in the  NE  region.  For  the
latter, data  for selected  nuclei, nuclear  matter and  in fact  for all
investigated  targets  treated  simultaneously, showed  universal  coarse
scaling  in   $y_G$.   The   above  then  also   holds  for   the  ratios
$\xi^{A_1,A_2}(y_G<0,Q^2)$.    The  same   results   \cite{don}  when   a
relativistic Fermi gas scaling variable is used \cite{bar}.

We report below on  such an analysis of the new CEBAF  data for the pairs
C,Fe and Fe,Au.  A novel element is the extension of the $y_G$-range into
the entire NI range $y_G\gtrsim 0$.   Although for all $\theta$ or $Q^2$,
cross sections per  nucleon for given $y_G$ span 4-5  decades (or about 3
decades for $\xi^{A,<N>}),\,\, \xi^{A_1,A_2}\approx 1$ within 15-20\% and
frequently  better  (see Table  I).   This  agrees with  the  approximate
$A$-independence (\ref{a3}) of nuclear  SF.  Occasional larger deviations
from 1 are ascribed to experimental uncertainties, in particular for data
of lowest intensity.

Next we consider  the EMC ratio $\xi^{A,N}_{NE}$, which is  not a special
case  of  $\xi^{A_1,A_2}$  with $A_2\to\langle  N\rangle=D/2$.   Density,
momentum distribution  and pair-distribution function for  $D$ differ and
$A  \gtrsim  12$.  Although  the  NE  and  the  NI regions  both  contain
information on the single-nucleon  momentum distribution $n(p)$, implicit
in $f^{PN}$, Eq. (\ref{a2}), simple  expressions for $\xi^{A,N}$ can only
be given for $y_G<0$.  However,  that portion is not directly observable.
In particular for high $Q^2$, NI  contributions compete with the NE ones,
even on  the elastic  side $y_G\lesssim 0$  ($x \ge 1$)  of the  QEP (cf.
Ref.  \onlinecite{rt2}, Fig. 4), as  has already  been realized  in the
analsis of early high-$E$ experiments on  the lightest nuclei D, $^3$ He,
$^4$He$\,$ \cite{day1}.

In  order   to  isolate   the  NE   part  of $\xi^{A,N}(y_G<0,Q^2) $,
one  has to remove  NI parts from  the data, which  can be
done in several  ways.  For instance (\ref{a2}) in  conjunction with (1),
provides a model for the NI parts, which gives perfect agreement with the
data for $y_G\ge  0$.  Although this need not be the same for $y_G\le 0$,
we assume this  to be  the  case.  The  procedure ultimately becomes
impractical,  because both  total and  NI parts  rapidly decrease  with
$y_G$  beyond $y_G\approx  -0.25$.   One  has then  to  rely on  directly
calculated  NE parts,  using  again  (\ref{a2}), now  with  NE parts  for
$F_k^N$.  The result is
\begin{mathletters}
\label{a5}
\begin{eqnarray}
F_1^{N(NE)}(x,Q^2)&=&\frac{x}{2}[G^N_M(Q^2)]^2\delta(x-1)
\nonumber\\
F_2^{N(NE)}(x,Q^2)&=&\frac{[G^N_E(Q^2)]^2+\eta[G_M^N(Q^2)]^2}{1+\eta}
\delta(x-1)
\label{a5a}\\
F_1^{A(NE)}(x,Q^2)&=&\frac{1}{2}f^{PN}(x,Q^2)[G^N_M(Q^2)]^2
\nonumber\\
F_2^{A(NE)}(x,Q^2)&=&xf^{PN}(x,Q^2)
\frac{[G^N_E(Q^2)]^2+\eta[G_M^N(Q^2)]^2}{1+\eta}
\label{a5b}
\end{eqnarray}
\end{mathletters}
Substitution in (1) yields for $\xi^{A,N}$, Eq. (\ref{a4})
($\eta=Q^2/4M^2$)
\begin{mathletters}
\label{a6}
\begin{eqnarray}
\xi^{A,N}_{NE}(x,Q^2)=f^{PN}(x,Q^2)
&&\bigg \lbrack \frac
{(x^2m^2/Q^2)([G^N_E]^2+\eta [G_M^N]^2)/(1+\eta)
+{\rm tan}^2(\theta/2) ([G_M^N]^2}
{(m^2/Q^2)[G^N_E]^2+\eta[G_M^N]^2)/(1+\eta)
+{\rm tan}^2(\theta/2)[G_M^N]^2}\bigg \rbrack
\label{a6a}\\
\xi^{A,N}_{NE}(x\approx 1,Q^2)&=&f{PN}(x\approx 1,Q^2)
\label{a6b}
\end{eqnarray}
\end{mathletters}
where for  better readability, we  have dropped the arguments  on $G^N$.

In  Fig. 2  we plot  $\xi_{NE}^{Fe,N}(y_G<0,Q^2)$ against  $Q^2$ for  a
number  of narrowly  binned $y_G$.   Whenever possible,  we give  the two
results, which with  the exception of $y_G=0$,  approximately agree.  For
all  $y_G\le 0$  in the  kinematic  range of  the experiment,  $\xi_{NE}$
approaches  the  AL  $Q^2\to\infty$  from  above,  or  can confidently be
extrapolated  to $Q^2$  beyond the  observed ones.   For the
largest $|y_G| $, there is hardly any $Q^2$ dependence.

We mention here an analysis of a PWIA scaling function which differs from
the  above $\xi^{A,N}$ primarily by a kinematic factor and by the use of
a different   scaling   variable $y_0=y^{PWIA}\,\,\,$  \cite{arr,ciof1}.
Arrington  $et\,  al$ found  that  for  $y_0\le -0.3$  GeV,  $\xi^{PWIA}$
approaches a  plateau from  above, with  persistent structures  for large
$Q^2$. However, for  $y_0=0.0,  -0.1$, $\,\,\xi^{PWIA}$  increases
with growing  $Q^2$.  Assuming that  NI contributions have  been properly
removed, the spelled-out  differences may be due to the  use of different
scaling variables.  Those  amount to the implicit  retention of different
FSI parts which are not easily  compared \cite{rj}.  Since the authors of
Ref. \onlinecite{arr}  plan studies in  still different variables,  we do
not pursue the issue here.

A plateau is  conventionally related to the AL, from which
one extracts  the single-nucleon momentum distribution  $n(p)$.  However,
the standard argument  becomes invalid, if parts of the  FSI happen to be
weakly $Q^2$-dependent,  causing the  plateau to  also contain FSI parts.
There are strong indications that this is approximately the case for
the kinematic  range of the  CEBAF experiment \cite{rt4}, which  seems to
contradict the behavior of $\xi^{A,N}$  for low $Q^2$, seen  in Fig. 1.

We give here  only  an  outline   of  the  resolution  of  the  apparent
contradiction:   details  may   be  found   in  Ref.   \onlinecite{rt2}.
The argument follows the construction of $f^{PN}(x,Q^2)$ from
the NR SF $\phi(q,y_G)$.  For smooth underlying $NN$ interactions
the $q$-dependence of FSI may as follows be written and parametrized
\begin{mathletters}
\begin{eqnarray}
\label{a7}
\phi(q,y)&=&F_0(y)+\phi^{odd}(q,y)+\Delta^{even}\phi(q,y)
\label{a7a}\\
F_0(y)=\lim_{q\to \infty}\phi(q,y)
&=&\frac{1}{4\pi^2}\int_{|y|}^{\infty} dp p n(p,\gamma_k),
\label{a7b}\\
\phi^{odd}(q,y)&=&U^{(o)}(q)
y\sum_{n}a_n^{(o)}y^{2n}{\rm exp}(-[A^{(o)}y]^2)
\label{a7c}\\
\phi^{even}(q,y)&=&U^{(e)}(q)
\sum_{n}a_n^{(e)}y^{2n}{\rm exp}(-[A^{(e)}y]^2)
\label{a7d}
\end{eqnarray}
\end{mathletters}
$F_0$ and $\Delta\phi$ are parts even in $y$, $\phi^{odd}$ is odd.

The notion of a
local, energy-independent $V_{NN}$ breaks down for increasing
lab momentum $q$, and $V$ has to be replaced $V\to V_{eff}=t_q$, the
latter being the off-shell elastic $T$ matrix for small momentum
transfer. It is the dominant ${\rm Im}[t(q,Q^2=0)]
\propto \sigma_q^{NN,tot}$ which is hardly $q$-dependent in the relevant
range. The link of $\phi$ with $f^{PN}$ in (\ref{a2}) reads \cite{rt2}
\begin{eqnarray}
f^{PN}(x,Q^2)&=&MD(x,Q^2)\phi\bigg ( q(x,Q^2),y_G(x,Q^2)\bigg ),\,\,
\nonumber\\
{\lim_ {Q^2 \gg4M^2x^2}}&& D(x,Q^2)=1,
\label{a8}
\end{eqnarray}
with $D$  a purely kinematic factor.
From (\ref{a6b}) one concludes that for $y\approx 0\,$ $(x\gtrsim 1$)
\begin{eqnarray}
\xi^{A,N}_{NE}(y_G\lesssim 0,Q^2)\approx f^{PN}(y_G\lesssim 0,Q^2)
=MD\phi(q,y_G\lesssim 0)
\label{a9}
\end{eqnarray}
Consequently
a weakly $q$-dependent $\phi(q,y_G)$ acquires $Q^2$ dependence through
$D$ for relatively low $Q^2$, which gets
weaker for increasing $Q$, as indeed observed.

It is of  interest to recall here  a suggested parametrization of
$\xi^{IA}(y_o,Q^2)$, function of $y_0$ \cite{atti}
\begin{eqnarray}
\xi^{IA}(y_0)=
\frac{C_1{\rm exp}(-a^2y_0^2)}{\alpha^2+y_0^2}+ C_2{\rm exp}(-b|y_0|)
\label{a10}
\end{eqnarray}
The above appears to provide a universal fit for $q$-dependent data, with
practically  $A$-independent parameters  for $A\gtrsim  12$,
which follows from Eq.  (\ref{a3}).  Moreover  we  note  that  an
expansion of (\ref{a10}) in $y_0$ leads to a form similar to (\ref{a7d}).
Without entering an  interpretation of the parameters,  this implies that
either all the data are in the AL  region, or that FSI in addition to the
AL are hardly $q$-dependent. If the latter is the case, the extraction of
$n(p)$ from the AL, without some knowledge of dynamics, becomes close  to
impossible.

Summarizing we  put into  focus the relation (\ref{a2}) which enables the
determination  of  nuclear SF  $F_k^A$  for  computed  SF $f^{PN}$  of  a
nucleus,  composed  of  point  nucleons.  The  latter  contains  the  AL,
expressed  by the  single-nucleon momentum  distribution $n(p)$,  and FSI
terms for finite $Q^2$.  Those  FSI, induced by binary collisions between
the struck  and a  core nucleon, depend  on effective  $NN$ interactions,
matter properties, none of which are known with great precision.

A first application is for inclusive cross and the agreement with data is
excellent for NI parts and also for the immediate NE side $x\gtrsim 1$ of
the QEP.  It  deteriorates for the most elastic regions  $x\ge 1$ for all
angles  where  cross  sections  are  smallest.   We  found  that  ternary
collision  contributions  do not  account  for  the above  discrepancies.
However, different single-nucleon momentum distributions easily cover the
range of the problematic parts of the cross sections.

We  performed a  scaling  analysis for  ratios $\xi^{A_1,A_2}(x,Q^2)$  of
cross  sections   for  different   nuclei  and  observed   the  predicted
$A$-independence  over the  entire NE  and  NI regions.   A second  ratio
$\xi^{A,N}(y_G,Q^2)$ was found to approach a plateau for all $y_G<0$ with
negative  slope, decreasing  for increasing  $Q^2$.  We  also
studied FSI  and
concluded  that those  are only  weakly $q$-dependent. Consequently the
plateau contains  the asymptotic limit  as well  as FSI.  The  same seems
also to be compatible with results  of a direct fit of scaling functions.
Should  the above  be confirmed,  there may  be less  incentive to  study
nuclear scaling  in the high-$Q^2$  regime as  a tool to  extract nucleon
momentum  distributions.   From  the agreement  between  predictions  for
$d^2\sigma$ and  data (Fig.  1) we conclude  that a  weakly $q$-dependent
FSI, which does not contradict any principle, is empirically correct.

In our conclusion we assemble results obtained for a variety of inclusive
scattering observables, which cover a wide range of kinematic ranges, and
which all are  related to computed nuclear form factors.   In addition to
the above mentioned  inclusive cross sections, we mention  $R$ ratios and
ratios  of  moments of  $F_k^A(x,Q^2)$  \cite{rt5}.   Generally there  is
agreement with  data, which  frequently span  broad ranges  of variation.
All prediction  use the underlying relation  (\ref{a2}) between structure
functions of  a nucleus and of  nucleons.  Of course, the  agreement does
not prove the  relation but neither does it disclose  definite flaws.  In
fact,  one wonders  whether (\ref{a2})  is the  result of  some effective
theory which, at least for  the chosen observables, successfully replaces
the fundamental QCD.

\bigskip

The authors thank John Arrington for having communicated data  and
Brad Fillipone for discussions and correspondence on the subject matter.

{Figure captions}

Fig. 1
Data \cite{arr} and predictions \cite{rt1} for the CEBAF 89-001
experiment.

Fig. 2.
The NE part of the GRS-type  scaling function
$\xi^{Fe,N(NE)}(y_G \le 0,Q^2)$ (\ref{a6a})
as function of $Q^2$. Dots connect extracted values, drawn
lines data, with calculated NI parts removed.

\newpage
%\makebox[8.5in]{\hfill{\bf Table II}\hfill}
%\begin{minipage}[b]{8.5in}
\begin{center}
{\bf Table I}
\vskip 1cm
%\vspace{1cm}
\begin{tabular}{|c|c||c|c||c|c|}
\hline
$\langle y_G\rangle$ (GeV)   &   $\theta$   & $x$  & $Q^2$ (GeV$^2$)  &
$\xi^{C,Fe}$ &$\xi^{Fe,Au}$ \\
\hline
\hline
     &  23 & 2.30 &     2.26    &  0.81 & 1.03   \\
-0.4 &  30 & 1.95 &     3.38    &  0.70 & 0.84   \\
     &  45 & 1.67 &     5.46    &  0.97 &  -     \\
\hline
     &  15 & 2.49 &     1.05    &  0.82 & 1.00   \\
-0.2 &  30 & 1.37 &     3.09    &  0.98 & 1.19   \\
     &  55 & 1.30 &     5.78    &  0.87 & 1.24   \\
\hline
     &  15 & 1.02 &     0.97    &  1.18 & 1.05    \\
 0.0 &  30 & 1.01 &     2.79    &  1.04 & 1.16    \\
     &  74 & 1.01 &     5.77    &  1.28 & 0.84    \\
\hline
     &  15 & 0.65 &     0.91    &  0.97 & 1.02    \\
 0.2 &  30 & 0.72 &     2.43    &  1.00 & 1.10    \\
     &  74 & 0.74 &     4.54    &  1.08 &   -     \\
\hline
 0.4 &  15 & 0.43 &     0.83    &  1.00 & 1.03    \\
\hline
\end{tabular}
\end{center}
%\vspace{1cm}
\vskip 1cm
Selection of  cross section ratios $\xi^{A_1,A_2}$,
Eq. (\ref{a4}).  For each selected, narrowly-binned $\langle y_G\rangle$,
available data of the ratios are given for
the smallest, some medium and largest ($x, Q^2$) in the data sets.
\newpage


\begin{references}

\bibitem{arr}
J.R. Arrington et al, Phys. Rev. Lett 82, 2056 (1999).

\bibitem{day}
D.B. Day $et\, al$, Phys. Rev. C{\bf 48}, 1849 (1993).

\bibitem{brad}
B.W. Fillipone, $et\,al$, Phys. Rev. C{\bf 45}, 1582 (1992).

\bibitem{arr1}
J. Arrington $et\, al$, Phys Rev. C{\bf 53}, 2248 (1996).

\bibitem{rt2}
A.S. Rinat and M.F. Taragin, Nucl. Phys. A {\bf 620}, 417 (1997);
Erratum: Nucl. Phys. {\bf A623}, 773 (1997).

\bibitem{gr}
S.A. Gurvitz and A.S. Rinat, TR-PR-93-77/
WIS-93/97/Oct-PH; Progress in Nuclear and Particle Physics,
{\bf 34}, 235 (1995).

\bibitem{bod}
A. Bodek and J. Ritchie, Phys. Rev. D23, 1070 (1981);
P. Amadrauz $et\, al$, Phys Lett. B295, 159 (1992); M. Arneodo $et\,al$,
$ibid$ B364, 107 (1995).

\bibitem{grs}
H.A. Gersch, L.J. Rodriguez and Phil N. Smith,
Phys. Rev. A{\bf 5}, 1547 (1973); H.A. Gersch and L.J. Rodriguez,
Phys. Rev. A{\bf 8}, 905 (1973).

\bibitem{aj}
S.V. Akulinitchev et al, Phys. Lett. B{\bf 158}, 485 (1985); Phys. Rev.
Lett. {\bf 59}, 2239 (1985); Phys. Rev. C{\bf 33}, 1551, (1986).

\bibitem{jaffe}
R.J. Jaffe, Nucl. Phys. A{\bf 478}, 3c (1988).

\bibitem{rt3}
A.S. Rinat and M.F. Taragin, Phys. Rev. B{\bf 58}, 15011 (1998).

\bibitem{om}
O. Benhar, A. Fabricioni, S. Fantoni, G.A. Miller,
V.R. Pandharipande and I. Sick, Phys. Rev. C{\bf 44}, 2328 (1991);
Phys. Lett. B{\bf 359}, 8 (1995).

\bibitem{ciof1}
C. Ciofi degli Atti, E. Pace amd G. Salm$\grave e$, Phys. Rev. C{\bf 43},
1155 (1991).

\bibitem{ciof2}
C. Ciofi degli Atti, D.B. Day and S. Liuti, Phys. Rev.
C{\bf 46}, 1045 (1994).

\bibitem{oset}
P. Fernandez de Cordoba, E. Marco, H. Mutter, E. Oset and A. Faessler,
Nucl. Phys. A{\bf 611}, 514 (1996).

\bibitem{ciof3}
C. Ciofi degli Atti and S. Simula, Phys. Lett.
B{\bf 325}, 276 (1994).

\bibitem{rj}
A.S. Rinat and B.K. Jennings, Phys. Rev. C{\bf 59} to be published.

\bibitem{rt1}
A.S. Rinat and M.F. Taragin, Nucl. Phys. A{\bf 598}, 349 (1996).

\bibitem{sag}
S.A. Gurvitz, Phys. Rev. C{\bf 42} (1990) 2653.

\bibitem{don}
T.W. Donnely and Ingo Sick, preprint nucl-th/9809063.

\bibitem{bar}
M.B. Barbaro, R. Cenni, A. dePace, T.W. Donnely and A. Molinari, Nucl.
Phys. A, to be published.

\bibitem{day1}
D.B. Day, J.S. McCarthy, T.W. Donnely and I. Sick, Ann. Rev. of Nucl. and
Particle Physics,  40, 357 (1990).

\bibitem{rt4}
A.S. Rinat and M.F. Taragin, submitted to Phys. Lett. B

\bibitem{atti}
C. Ciofi degli Atti, D. Faralli and G.B. West, Elba Workshop on Electron-
Nucleus scattering, EIPC, June 1998.

\bibitem{rt5}
A.S. Rinat and M.F. Taragin, submitted to Phys. Rev. C.

\end{references}
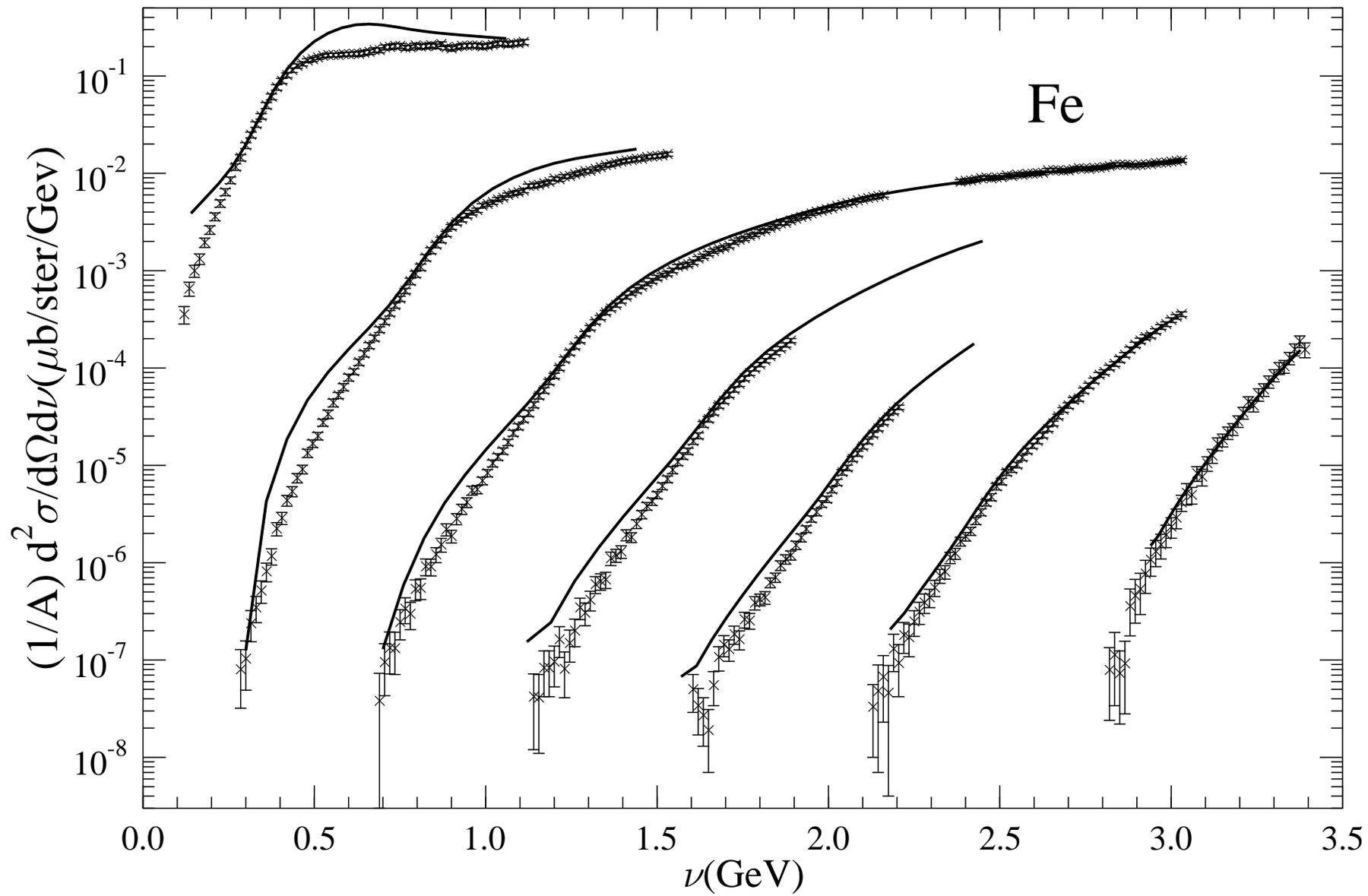
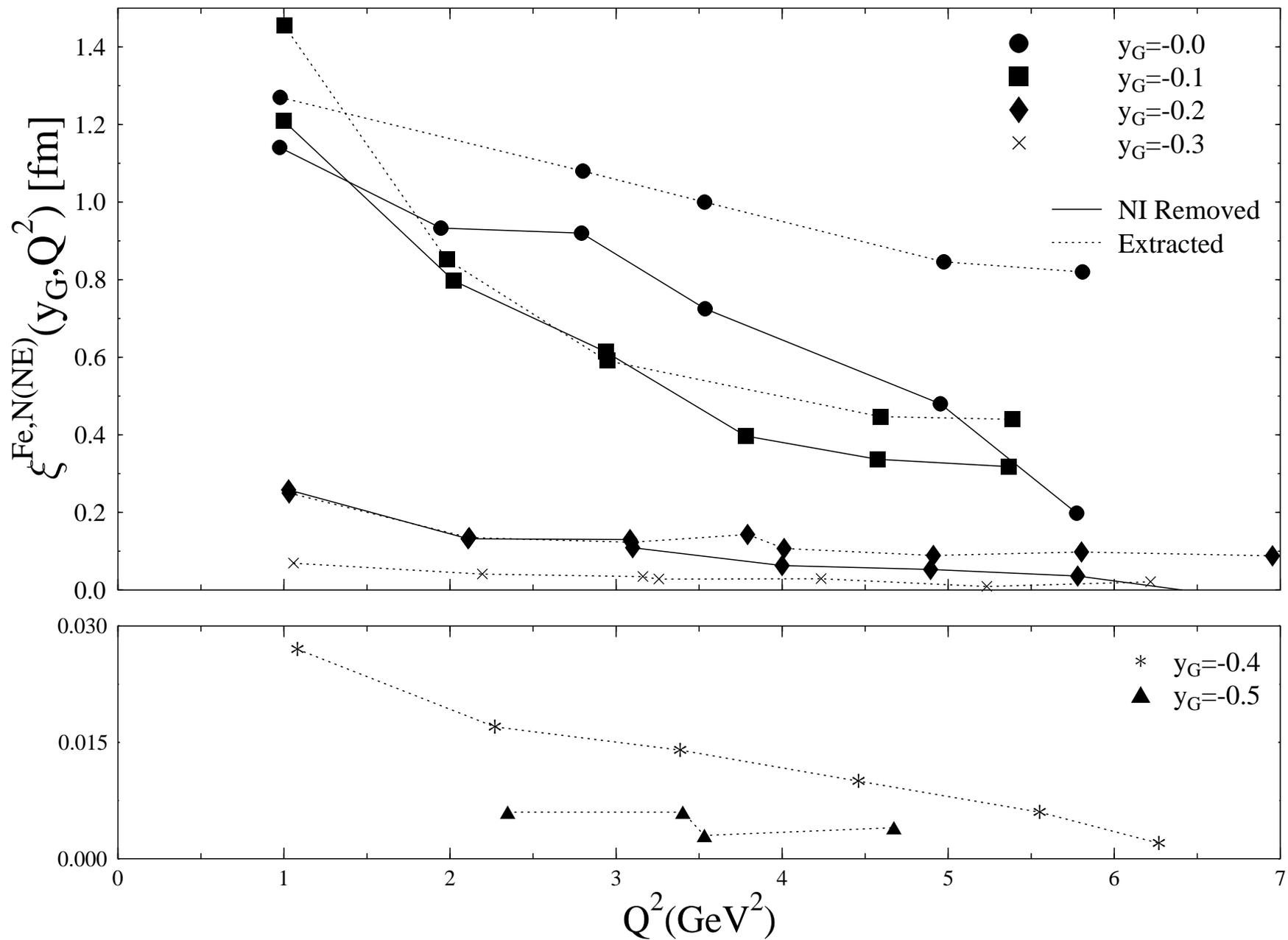
\end{document}